\documentclass[aps,prd,superscriptaddress,floatfix]{revtex4-1}
\usepackage{color}
\usepackage{amssymb}
\usepackage[normalem]{ulem}
\usepackage{graphicx,color}
\usepackage{amsmath}
\usepackage{subfigure}
\usepackage[normalem]{ulem}
\usepackage{booktabs}
\usepackage{xcolor}
\usepackage{epsfig,graphics,color,graphicx,amsmath}

\colorlet{darkgreen}{green!50!black}
\colorlet{brightyellow}{yellow!75!red}
\colorlet{orange}{red!50!yellow}
\colorlet{darkblue}{blue!60!black}
\colorlet{darkred}{red!80!black}

\usepackage{soul}

\usepackage{mathtools}
\usepackage{mathrsfs}
\usepackage{bm}
\usepackage{relsize}
\usepackage{indentfirst}

\usepackage{xcolor}

\usepackage{amssymb,amsmath,multirow,epsfig,graphicx,color}

\usepackage{epsfig}
\usepackage{graphicx,amsmath}
\def\be{\begin{eqnarray} &&}

\def\ee{\end{eqnarray}}

\newcommand\ba{\begin{eqnarray}}
\newcommand\ea{\end{eqnarray}}

\newcommand{\bas}{\begin{eqnarray*}}
\newcommand{\eas}{\end{eqnarray*}}

\newcommand{\bno}{\begin{eqnarray*}}
\newcommand{\eno}{\end{eqnarray*}}

\def\sl
 \usepackage{hyperref}
\usepackage{url}
 \usepackage{hyperref}
\hypersetup{
	colorlinks=true,  
	linkcolor=blue,   
	filecolor=magenta,      
	urlcolor=blue,    
	citecolor=blue,   
} 
\begin{document}
\vspace{-12ex}    
\begin{flushright} 	
{\normalsize \bf \hspace{50ex}}
\end{flushright}
\vspace{11ex}
	 \title{Void-dominated Cosmology: Cosmological Constant Problem and Hubble Tension}
\author{E. Yusofi}
\email{eb.yusofi@iau.ac.ir(Corresponding author)}
\affiliation{Department of Physics, Ayatollah Amoli Branch, Islamic Azad University, Amol, Iran}
\affiliation{School of Astronomy, Institute for Research in Fundamental Sciences(IPM), P. O. Box 19395-5531,Tehran, Iran}
\affiliation{Innovation and Management Research Center, Ayatollah Amoli Branch, Islamic Azad University, Amol, Mazandaran, Iran}

\author{M. A. Ramzanpour}
\email{ma.ramzanpour@iau.ac.ir}
\affiliation{Department of Physics, Ayatollah Amoli Branch, Islamic Azad University, Amol, Iran}
\affiliation{Innovation and Management Research Center, Ayatollah Amoli Branch, Islamic Azad University, Amol, Mazandaran, Iran}

\date{\today}
\begin{abstract}
In this work, we study the cosmological constant problem and Hubble tension in the void-dominated cosmology scenario \cite{Yusofi:2022hgg}. For this goal, we will first consider the cosmic voids in the cosmic web as interconnected ideal spherical bubbles at the early Planck scale and late large scale. By heuristic calculations for each cosmic void, we obtain particular mass density and cosmological constant that are the same order of magnitude as the entire universe on any scale. The values obtained for the cosmological constant vary from scale to scale. As a result, it will be shown that there is a roughly $\sim 10^{22}$ difference between the cosmological constant at the present and Planck scales. Finally, it will be shown that the slight difference between the surface tension of the cosmic bubbles may explain the tension between local and global measurements of $H_{\rm 0}$.  \\
\noindent \hspace{0.35cm} 
\noindent \hspace{0.35cm} 
\end{abstract}

\maketitle
\section{Introduction and Motivation}\par Some of major problems in modern cosmology are dark matter, dark energy, cosmological constant, and Hubble tension \cite{Silk:2016srn, Peracaula:2022vpx,Abdalla:2022yfr}. These serious physical challenges show that the standard model of cosmology, with all its advantages, is not able to solve many of these problems accurately\cite{Perivolaropoulos:2021jda, Abdalla:2022yfr}. Dark energy is an energy that the cause of the accelerating expansion of the present universe at large scales. In \cite{Belgacem:2021ieb} has been considered dark energy originates from the amplification of quantum fluctuations of a light field during inflation. Also, dark matter as the invisible and ghost-like matter is the cause of the, interconnection and balance of the galaxies, clusters and superclusters. Both of dark matter and dark energy has been completely confirmed by various direct and indirect observational methods. But for decades no plausible physical source has been found for this major contribution to the universe's matter and energy \cite{Zwicky:1937dam, Peebles:2002gy}.\\
When the Hubble parameter is measured in local scales by cosmic ladders scaling methods, the values measured for $H_{\rm 0}$ are different from the value reported by Planck, which depends standard $\Lambda$CDM model and uses CMB photons. This difference is called the \textit{Hubble tension}. On the one hand, scientists are looking for more accurate scaling methods, and on the other hand, they are looking for an alternative to the standard model for solving Hubble tension\cite{Jang:2017dxn,Vagnozzi:2019ezj,Haslbauer:2020xaa,Riess:2020fzl,Pesce:2020xfe,Kim:2020gai,Abadi:2020hbr,Moshafi:2020rkq,Krishnan:2020vaf,Vagnozzi:2021gjh,DiValentino:2021izs, Odintsov:2022eqm, Belgacem:2021ieb}.\\
In the structure of the cosmic web, the cosmic voids occupy the main volume of empty space and the shells of these interconnected voids act as the scaffolding of this huge structure. Although the size of the cosmic voids is much larger than the size of local scales and much smaller than the size of cosmic scale, it will be shown that they can be studied as a cell of the cellular structure of the cosmic web (foam). For the first time by considering a void-dominated cosmology \cite{Yusofi:2022hgg}, the surface tension of cosmic (Vast/Super) voids is introduced as a possible physical source for dark matter and dark energy \cite{Yusofi:2019sai,Yusofi:2022hgg}. We have assumed that the supervoids dominate in the in the large-scale overviwe. The gravitational integration of galaxies over time, on the one hand, leads to the formation of over-dense regions such as clusters, superclusters, walls, strings, filaments, and nodes. On the other hand, as superclusters merge, almost empty spaces are created between them, and we call these under-dense regions among galactic strings and walls as cosmic voids\cite{Weygaert:2011}. Hydrodynamic models and simulations of the formation of the cosmic web structure show that these voids are also merging\cite{Carlesi:2014kua}.\\
In the standard model of cosmology ignores the statics, dynamics and evolution of supervoids that make up the main volume of the vacuum-like fluids of early and late cosmos. While the supervoids are not only not completely empty, but also have an energy density and evolution, and also because they are bulky, they are more likely to merge with each other and are much more suitable candidates for influence on the cosmic scales\cite{Weygaert:2011, Higuchi:2017hdo, Yusofi:2022hgg,Pandey:2019qcb,Pandey:2018fkb}. Therefore, the possibility of the role of these large inhomogeneities in the dynamics of the universe and the effects of their evolution in determining the value of cosmic parameters seems possible\cite{Nan:2021prt,Vigneron:2019dpj,Buchert:2018vqi,Srivastava:2007en}.\\
In the void-dominated cosmology \cite{Yusofi:2022hgg}, voids (bubbles) constitute the main distribution of the cosmic two-phase fluid in both the early inflationary period and the current accelerated universe \cite{Yusofi:2018lqb}. In the proposed hypothesis in \cite{Yusofi:2022hgg}, cosmic voids are assumed to be spherical bubbles their shells are surrounded by galactic superclusters. By considering the shells as the ideal separating surface between the under-density bubble-like areas and the over-density droplet-like areas, we have obtained the resulting surface tension by dimensional and heuristic calculation (see Table \ref{table:I})\cite{Yusofi:2022hgg}. Then, by equating the energy density of cosmic voids with the vacuum energy density, we show that the value estimated for the cosmological constant is very close to that predicted by Planck's observations and has the same order of magnitude \cite{Yusofi:2022hgg}. Our hypothesis about the effect of supervoids on the large-scale dynamics of the universe and cosmology  in \cite{Yusofi:2022hgg} could be consistent with the backreaction effects of density inhomogeneities in cosmology \cite{Buchert:2015iva} and the configurational entropy of the cosmic web and
its evolution \cite{Pandey:2019qcb,Pandey:2018fkb}. But it can be inconsistent with the Green and Wald results \cite{Green:2016cwo}. \\
In this paper, in Sec. II, we discuss the simultaneous coexistence of super-voids and super-clusters as two evolving part in the Cosmic Web. In the void-dominant universe by heuristic calculations we obtain particular mass densities, surface tensions and cosmological constants for each void in Sec. III. and compare its magnitude with the density of the whole universe. We will then try to answer two important following questions:\\
\textbf{i.} Given that the average diameter of cosmic voids is usually $100  {\rm Mpc}$, might a single cosmic void be a good representation for smaller local scales and larger global scale?\\ \textbf{ii.} Do the slightly differences between of the surface tension of different voids obtained from our hypothesis can represent the acceptable values of the Hubble constant? and could the discrepancy between their values be a possible explain for $H_{\rm 0}$ tension?\\
In the final section of this paper, we will briefly discuss the possibility of resolving $H_{\rm 0}$ tension and other possible outcomes.
\begin{table}
	\caption{The surface tension of $ \gamma_{\rm i} $, on the shell of cosmic voids containing disk-shaped superclusters $i= 1, 2, 3, 4 $ \cite{Yusofi:2022hgg}.}
	
	\begin{center}
		\begin{tabular}{c c c c}
			\hline
			i.  Largest&\quad $M_{\rm i}$&\quad $R_{\rm i}$&\quad \quad $\gamma_{\rm i}$ \\
			\quad scale\quad&\quad($10^{47}{\rm {kg}}$)\quad & \quad($10^{24}{\rm {m}}$) \quad&\quad $(10^{15}{\rm {J.m^{-2}}})$ \\
			\noalign{\smallskip}\hline
			\noalign{\smallskip}\hline
			1. Corona Sc &\quad $0.20$ &\quad $1.50$ &\quad $0.25$ \\
			2. Virgo Sc &\quad $0.03$ &\quad $0.50$ &\quad $0.34$\\
			3. Laniakea Sc &\quad $1.00$ &\quad $2.40$ &\quad $0.50$\\
			4. Caelum Sc &\quad $4.00$ &\quad $ 4.30$ &\quad $0.62$ \\
			\noalign{\smallskip}\hline
	    	\end{tabular}
	\end{center}
	\label{table:I}
\end{table}

\section{Coexistence of Supervoids and Superclusters in Cosmic Web}
\par If we consider part of the present cosmos, it contains a network of cosmic voids in which several superclusters and small and large galactic objects are merging with each other. It is certain that in the global view, the large scale of the universe is in the void-dominant situation, but in the small over-dense regions, it includes clusters, filaments, and nodes of normal matter situated in the matter-dominant situation. The coexistence and continuous integration of superclusters and merging of supervoids increase the contribution of each of them to the structure of the universe, and the increase in the size of the cosmic void after merging them leads to an effective repulsive force on the galaxies situated on their shell \cite{Weygaert:2011, Yusofi:2022hgg}. Under such conditions, it can be assumed that the cosmic fluid at the large-scale overview consists of merging and expanding large voids, and the universe is dominated over time by larger cosmic voids and as a result, we can have an accelerating expansion universe.\\
In the proposed model, over-dense objects, including galaxies and their clusters and superclusters, are thought of as "drops" and the under-dense voids, and supervoids between them as "bubbles"\cite{Yusofi:2022hgg, Yusofi_2010}. In a two-phase inhomogeneous mixture of drops and bubbles, the bubbles also absorb other bubbles and disperse the droplets from the center to the boundary, while their size becomes larger than before. Physical simulations of redshifts in terms of different displacements show that local scales become denser over time but the density of large scales decreases\cite{Weygaert:2011}.\\

\section{Global and Local Behavior of Voids in Void-dominated Cosmology}
\par In this part of the paper, we want to show whether a single cosmic void can be a good representation of local and global scales or not? For this purpose, we first calculate the mass density of an ideal spherical supervoid and show that its magnitude is about one-tenth of the average density of the whole universe. Then, considering the surface tension of a cosmic bubble, we show that the cosmological constant obtained from it is very close to the cosmological constant measured by Planck 2018.\cite{Yusofi:2022hgg, Planck:2018jri}. Finally, we will calculate the cosmological constant at very early Planck scale and will show that its value is $\sim 10^{22}$ times the observed value.

\subsection{Mass Density for a Cosmic Void}
For a perfectly empty spherical bubble with a total mass accumulated on the shell, the mass density can be calculated from the simple relation below,
\begin{equation}
\label{hob1}
\rho_{\rm i}=\frac{3{M_{\rm {i}}}}{4\pi {\bar{r}_{\rm {v}}^3}}.
\end{equation}
Here $M_{\rm {i}}$ is the mass of the supercluster enclosed the cosmic void and $\bar{r}_{\rm {v}}$ is the average radius of a cosmic void. Taking into account the values of Table \ref{table:I}. for the mass and radius of the Laniakia supercluster, we obtain
\begin{equation}
\label{hob2}
\rho_3=1.70 \times 10^{-27} {\rm {kg.m^{-3}}}.
\end{equation}
is very close to the universe's average mass density of the universe \cite{Cheng:2008grc}\textit{i.e.}
\begin{equation}
\label{hob3}
\rho_{0,c}=1.88 \times 10^{-26} {\rm {kg.m^{-3}}}.
\end{equation}
 and is about one order smaller than that. This density deficit appears to be due to the assumption of a completely empty bubble, while inside the cosmic voids there is galactic gas with a lower density than the shell. The mass densities of the other cosmic voids are listed in Table \ref{table:II}, respectively.
\subsection{Cosmological Constant for a cosmic void}
The internal pressure of a single bubble (drop) is usually greater than its external pressure, and the pressure difference with the outside comes from the Young-Laplace formula \cite{Butt:2003pci,Reichl:2016msp}
\begin{equation}\label{hob4}
\Delta{P} = \frac{2\gamma}{\bar{r}_v}.
\end{equation}
Here, $\gamma$ represents the surface tension for bubble (drop). To calculate the surface tension of a single bubble, we use the following heuristic method \cite{Yusofi:2019sai,Yusofi:2022hgg},
\begin{equation}
\label{hob5}
\gamma_{\rm i}\equiv\frac{\rm Energy}{\rm Area}={\frac{M_{\rm i}c^2}{\pi R_{\rm i}^2}}.
\end{equation}
Taking account of the calculations in the previous section, the average density of a cosmic fluid is very close to the density of a single bubble. In the present void-dominant cosmic fluid we can assume that ($\rho_{\rm \Lambda} \equiv \rho_{\rm {v}}$) and ($\Delta{P} \simeq P_{\rm v}$), by considering (\ref{hob4}) we obtain
\begin{equation}
\label{hob6}
P_{\rm v} = w{c^2}\rho_{\rm v}  
\end{equation}
Therefore, to have a cosmological constant caused by bubbles, we will reach the following relation \cite{Yusofi:2022hgg},
\begin{equation}
\label{hob7}
\Lambda_{\rm i} =\frac{8\pi{G}}{{w{c^4}}}\frac{2\gamma_{\rm i}}{\bar r_{\rm {v}}}.
\end{equation}
By placing the necessary values \cite{Yusofi:2022hgg}, we reach the results in Table \ref{table:II}.\\
\subsubsection{The Smallest Cosmological Constant for Late Universe}
The cosmological constant in the latest Planck data is reported as below \cite{Planck:2018vyg},
\begin{equation}
\label{hob8}
\Lambda_{\rm {obs}}= 1.1056 \times 10^{-52} {\rm {m^{-2}}}.
\end{equation}
In our model, for the biggest object such as Laniakia supercluster in which the Milky Way galaxy is located inside of it, the cosmological constant of the model is obtained as follows\cite{Yusofi:2022hgg},
\begin{equation}
\label{hob9}
\Lambda_3 =  1.2979 \times  10^{-52} {\rm {m^{-2}}}.
\end{equation}
As we can see in Table \ref{table:II}, the cosmological constant and the mass density for each of the cosmic voids are the same as values for the entire universe (\ref{hob8}) and(\ref{hob3}) and are very close to them. Thus, given the values obtained for the $\rho_{\rm i}$ and $\Lambda_{\rm i}$, it seems that a cosmic void can be a good indicator of the global behavior of the accelerating universe. Therefore, the surface tension of cosmic voids may be regarded as a possible source producing dark energy\cite{Yusofi:2022hgg}.
\subsubsection{The Largest Cosmological Constant for Planck Scale}
In void-dominated cosmology, the cosmological constant has variable values on different scales. Since the sizes of voids are diverse, we must obtain the largest values for surface tension and cosmological constant in the smallest scales and vice versa. To show this issue more precisely, we consider the early foam-like space as very small Planck-sized identical bubbles connected to each other. Then, we calculate the surface tension and the cosmological constant for a hypothetical single bubble with a proton- and Planck-sized and present in Table \ref{table:III}. As a result, our classical model predicts a discrepancy between the cosmological constant values at the Planck scale and the present observable universe of the order of $\sim 10^{122}$ , which is consistent with the cosmological constant problem. 
\section{Is Hubble Tension from Bubble Tension?!}
For the small differences between the values of the cosmological constant for the different supervoids listed to Table \ref{table:II}, we will obtain slight discrapancy between the corresponding Hubble constants. We will show that our void-based model confirms $H_0$ values that reported locally measured \cite{Jang:2017dxn,Riess:2020fzl,Pesce:2020xfe} and the value inferred from the cosmic microwave background (CMB) \cite{Aloni:2021eaq}. Assuming the $\Lambda$CDM-based cosmology, the Hubble constant of the late universe (model dependent) is inferred as \cite{Planck:2018vyg},
\begin{equation}
\label{hob10}
H_{\rm {0_{global}}} = 67.66 \pm 0.42 \quad 
{\rm km.s^{-1}.Mpc^{-1}}
\end{equation}
 \begin{table}
	\caption{Mass density $\rho_{\rm i}$, cosmological constant $\Lambda_{\rm i}$, global and local Hubble constant ${H}_{\rm {0_{iG}}}$ and ${H}_{\rm {0_{iL}}}$ for different cosmic voids surrounded with superclusters $i= 1, 2, 3, 4 $, and average value of the parameters.}
	
	\begin{center}
		\begin{tabular}{c c c c c}
			
			\hline\noalign{\smallskip}
			Cosmological&\quad\quad $\rho_{\rm i}$&\quad\quad $\Lambda_{\rm i}$& ${H}_{\rm {0_{iG}}}$\quad \quad${H}_{\rm {0_{iL}}}$\\
			\quad Parameter&\quad($10^{-26}{\rm {kg.m^{-3}}}$) \quad&\quad $(10^{-52}{\rm {m^{-2}}})$\quad &\quad $({\rm {km.s^{-1}.Mpc^{-1}}})$ \\
			\noalign{\smallskip}\hline
			\noalign{\smallskip}\hline
			1. Corona Sc&\quad $0.14$ &\quad $0.6645$ &\quad $52.45$ \quad $57.39$  \\
			2. Virgo Sc&\quad $0.60$ &\quad $0.8970$ &\quad $60.94$ \quad $66.68$  \\
			3. Laniakea Sc &\quad $0.17$ &\quad $1.2979$ &\quad $73.31$ \quad $80.21$ \\
			4. Caelum Sc&\quad $0.12$ &\quad $1.6172$ &\quad $81.83$ \quad $89.53$\\
			\noalign{\smallskip}\hline
			Average value&\quad $0.25$ &\quad $1.1192$ &\quad $67.13$ \quad $73.45$\\
			\noalign{\smallskip}\hline
			\noalign{\smallskip}\hline
		\end{tabular}
	\end{center}
	\label{table:II}
\end{table}
\begin{table}
\caption{The surface tension of $ \gamma_{\rm i} $, and cosmological constant $ \Lambda_{\rm i}$ in Planck scale.}
	
	\begin{center}
		\begin{tabular}{c c c c c}
			\hline
			Smalest object&\quad $M_{\rm i}$&\quad $R_{\rm i}$& $\gamma_{\rm i}$ \qquad \qquad\qquad\qquad $\Lambda_{\rm i}$  \\
			\noalign{\smallskip}\hline
			\noalign{\smallskip}\hline
			Proton &\qquad $1.67\times10^{-27}{\rm kg}$& \qquad $1.00$$\times 10^{-15}{\rm m}$ &\qquad$4.7\times10^{19}{\rm j.m^{-2}}$ \qquad$1.94\times 10^{-8}{\rm m^{-2}}$\\
			Planck unit &\qquad $2.18\times10^{-8}{\rm kg}$& \qquad $1.62$$\times 10^{-35}{\rm m}$ &\qquad$2.38\times10^{78}{\rm j.m^{-2}}$ \qquad$6.08\times 10^{70}{\rm m^{-2}}$\\
			\noalign{\smallskip}\hline
			\end{tabular}
	\end{center}
	\label{table:III}
\end{table}
  . But from Hubble Space Telescope (HST) observations of 70 long-period Cepheids in the Large Magellanic Cloud, the best measurement of the cosmological constant has been estimated as \cite{Riess:2019cxk},
  \begin{equation}
  \label{hob11}
  H_{\rm {0_{local}}} = 74.03 \pm 1.42 \quad 
  {\rm km.s^{-1}.Mpc^{-1}}
  \end{equation}
 Now, we will continue to calculate the cosmological constant in our void-dominant model.\\
 The Hubble constant is related to the cosmological constant according to the following relation
  \begin{equation}
 \label{hob12}
 H_{\rm {0_i}}^2 =\frac{\Lambda_{\rm {i}}}{3\Omega_{\rm \Lambda}}c^2
 \end{equation}
 So $H_{\rm 0_i} \propto \Lambda_{\rm i}^{\frac{1}{2}}$ and we will have
  \begin{equation}
 \label{hob13}
 H_{\rm {0_i}} = H_{\rm {0}_{obs}}\left(\frac{\Lambda_{\rm i}}{\Lambda_{\rm {obs}}}\right)^{\frac{1}{2}}\end{equation}
 For observational Hubble constant $ H_{\rm {0}_{obs}}$, we consider two selections (\ref{hob10}) and (\ref{hob11}).\\
 If we consider $ H_{\rm {0_{obs}}}= 67.66\quad {\rm km.s^{-1}.Mpc^{-1}}$ from Planck 2018 data \cite{Planck:2018vyg} and considering observational cosmological constant (\ref{hob8}), the equation (\ref{hob13}) gives the following value for the cosmic void that surrounded by Laniakea supercluster
  \begin{equation}
 \label{hob14}
 H_{\rm {0_{3G}}} = 73.31 \quad 
 {\rm km.s^{-1}.Mpc^{-1}}
 \end{equation}
 On the other hand if we consider $ H_{\rm {0_{obs}}}= 74.03\quad {\rm km.s^{-1}.Mpc^{-1}}$, for the cosmic void that surrounded by Virgo supercluster we obtain
 \begin{equation}
 \label{hob15}
 H_{\rm {0_{2L}}} = 66.68 \quad 
 {\rm km.s^{-1}.Mpc^{-1}}
 \end{equation}
 The predicted value (\ref{hob15}) is very close to the value obtained in \cite{Kim:2020gai}.
 For other cosmic voids, the values of the Hubble constant are also listed in Table \ref{table:II}. Given the value obtained (\ref{hob14}) and (\ref{hob15}) in the proposed bubble model, it can be concluded that the Hubble constant values in it are  close to the values (\ref{hob10}) and (\ref{hob11}), in which measured by both of the global and local groups, respectively. As one can see in Table \ref{table:II}, the values obtained for the Hubble constant in our model include the values reported by both global and local measurements. \\
 However, given the value obtained for the Laniakea supercluster (\ref{hob14}), the model results are closer to local measurement(\ref{hob11}). Also, the average value of the Hubble constant of the void-dominated model can be obtained as follows \begin{equation}
 	\label{hob116}
 	{H_{\rm 0_{ave}}} = \frac{\Sigma H_{0_i}}{4}.
 \end{equation}
 For the average value of global $H_0$, we get $67.13\quad {\rm km.s^{-1}.Mpc^{-1}} $ and for local $73.45\quad {\rm km.s^{ . -1}.Mpc^{-1}} $, which are very close to the values reported from two observational groups i.e. (\ref{hob10}) and (\ref{hob11}).\\
 Given the high accuracy of measurements that reported by local groups on the one hand and the independence of these data from the model on the other, it seems that the main reason for $H_0$ tension is related to $\Lambda$ in the standard $\Lambda$CDM model that assumed \textit{completely constant}. Since, according to our hypothesis, the surface tension values of the supervoids are slightly different, as a consequent of it we can have different $H_{\rm 0}$. Thus, given the values obtained for the Hubble constant $H_{\rm i}$ in Table \ref{table:II}, it seems that cosmic voids can be a good indicator to study on both the global and local scales. It is shown that the KBC void and the Hubble tension are in conflict with the standard $\Lambda$CDM model at the Gpc scale \cite{Haslbauer:2020xaa}. Also, in \cite{Moshafi:2020rkq} has been shown that the observed differences in the measurements of Hubble parameter at local and global scales may indicate a need to modify $\Lambda$CDM model. Interestingly, in the modified ${\rm \ddot{u}}\Lambda$CDM model, $H_{\rm 0}$ values are closer to local measurements (see Table 3. in \cite{Moshafi:2020rkq}). Assuming the $H_{\rm 0}$ tension is cosmological in origin, in \cite{Krishnan:2020vaf} has been predicted that a running of $H_{\rm 0}$ with redshift can be expected. These additional determinations of $H_{\rm 0}$ may be traced to a difference between the effective EoS of the Universe within the standard model.\\
 Finally, the $H(z)$ values listed in Table 1. on page 5 of \cite{Gomez-Valent:2018hwc}, which have been published in various journals, show that the Hubble parameter decreases as the redshift $z$ decreases. This is a confirmation of our model that with the passage of time and as the radius of the cosmic voids increases, their surface tension decreases. Therefore, we now have the lowest possible value for the surface tension, cosmological constant, and the Hubble expansion as $ H(0)=H_0$ at $z=0$, which both the observations and the proposed model are completely consistent with it.
 \section{Discussions and Conclusions}
We have considered supervoids as ideal expanding spherical bubbles in a void-dominated cosmic fluid. The total supervoid mass is situated on the shell and the shell is formed by the disk-shaped superclusters. Then, by heuristic calculating the mass density and cosmological constant of a single supervoid, we have been shown that the value obtained is very interestingly the same as their corresponding values for the entire universe.\\
On the other hand, for the Hubble tension, we have been shown that the value obtained from the void-dominant universe hypothesis can represent the values obtained from local and global data, but is more consistent with local measurements. As we know, the data reported from global groups such as Planck depends on the $\Lambda$CDM model. Therefore, small changes in $\Lambda$'s value can greatly affect the results of Planck measurements. But on the other hand, the data reported from local groups are independent of each model and have a very high measurement accuracy. \textit{From our point of view, the main problem in Hubble tension may be originated from the slight changes in $\Lambda$'s value, which also depends on surface tension and the size of cosmic voids} in the structure of cosmic web ~\cite{Yusofi:2022hgg}. Highlights of this work include:\\
\\
1) The cosmic fluid/web consists of two simultaneous coexist dynamical parts, namely galaxies, clusters, superclusters, and nodes on one hand and small voids and supervoids on another hand.\\
\\
2) In void-dominated cosmology, voids are dominant and their existence and evolution play an essential role in the accelerating expansion of the cosmos.\\
\\
3) It has been shown that a cosmic void can be considered a very good representative for studying both local and global scales.\\
\\
4) For each cosmic void, we obtained a particular mass density and the cosmological constant that in terms of the order of magnitude are the same as the entire universe on the smallest and large scale. Also, it is shown that the largest discrepancy between the observable cosmological constant and the early foam-like cosmos ($\sim 10^{122}$) can be logical in the void-based model of the universe.
\\
\\
5) Finally, it has also been shown that the slight difference between the surface tension of the cosmic bubbles may be the source of tension between global and local measurements of $H_{\rm 0}$.\\
So the interesting result of this study is that a cosmic void as a cell of the cellular structure of the cosmic web scaffolding can be a good candidate to describe the behavior of the universe both on a global and local scale. \\
As a significant consequence of this research, by more exact probing the behavior of a single cosmic void more seriously, both theoretically and observationally, possible plausible solutions to the important challenges of physical cosmology on a local and global scales can be offered. In future works, we will address important issues such as dark matter and vacuum energy problems within the framework of void-dominated cosmology.

\section*{Declaration of Competing Interest}
The authors declare that they have no known competing financial interests or personal relationships that could have appeared to influence the work reported in this paper.
\section*{Acknowledgements}
EY would like to acknowledge Dr. A. Talebian for his constructive discussions and for his help in drawing the tables. This work has been supported by the Islamic Azad University, Ayatollah Amoli Branch, Amol, Iran.\\

\bibliography{VDM-CC-Bucalter.bib}

\end{document}